\documentclass[twocolumn]{article}

\ifdefined\nopdfsync\else\usepackage{pdfsync}\fi 

\advance\textheight by 3cm
\advance\topmargin -2cm
\advance\textwidth by 2.4cm
\advance\oddsidemargin by -1.2cm
\advance\evensidemargin by -1.2cm

\usepackage{amsmath}
\usepackage{amsfonts}
\usepackage{amssymb}
\usepackage{mathrsfs}
\usepackage{latexsym}
\usepackage{graphicx}
\usepackage{url}
\usepackage[show]{chato-notes}
\usepackage{algorithm, algorithmic}
\floatname{algorithm}{Algorithm}
\usepackage{booktabs}

\newcommand{\rst}[1]{\ensuremath{{\mathbin\upharpoonright}%
\raise-.5ex\hbox{$#1$}}}

\title{Layered Label Propagation: A MultiResolution Coordinate-Free
Ordering for Compressing Social Networks}
\author{Paolo Boldi\quad Marco Rosa\quad Massimo Santini\quad Sebastiano
Vigna \\Dipartimento di Scienze dell'Informazione, Universit\`a degli Studi di
	Milano, Italy}
	
\begin{document}
\bibliographystyle{alpha}
\maketitle

\begin{abstract}
We continue the line of research on graph compression
started in~\cite{BoVWFI}, but we move our focus to the compression of
\emph{social networks} in a proper sense (e.g., LiveJournal):
the approaches that have been used for a long time to compress web graphs
rely on a \emph{specific} ordering of the nodes (lexicographical
URL ordering) whose extension to general social networks is not trivial. In this
paper, we propose a solution that mixes clusterings and orders, 
and devise a new algorithm, called \emph{Layered Label Propagation}, that builds
on previous work on scalable clustering and can be used to reorder very large
graphs (billions of nodes). Our implementation uses \emph{task decomposition} to
perform aggressively on multi-core architecture, making it possible to reorder
graphs of more than 600 millions nodes in a few hours.
Experiments performed on a wide array of web graphs
and social networks show that combining the order produced by the proposed
algorithm with the WebGraph compression framework provides a major
increase in compression with respect to all currently known techniques, 
\emph{both} on web graphs \emph{and} on social networks.
These improvements make it possible to analyse in main memory significantly
larger graphs.
\end{abstract}

\section{Introduction}
\label{sec:intro}
The acquaintance structure underlying a social network contains a wealth of 
information about the network itself, and many data mining tasks can be 
accomplished from this information alone (e.g., detecting outlier nodes, 
identifying interest groups, estimating measures of centrality 
etc.~\cite{wasserman-socnet,Knoke2008}). 
Many of these tasks translate into graph mining problems and can be solved
through suitable (sometimes, variants of standard) graph algorithms that
often assume that the graph is stored into the main memory. However, this 
assumption is far from trivial when large graphs are dealt with, and this is
actually the case when social networks are considered; for instance,
current estimates say that the indexable web contains at least
$23.59$ billion pages\footnote{\texttt{http://www.worldwidewebsize.com/}}, and
in 2008 Google announced to have crawled 1 trillion unique URLs: the successor 
lists for such a graph would require hundreds of terabytes of memory! 
The situation is somewhat similar in other social networks; for
example, as of October 2010\footnote{\texttt{http://www.facebook.com/press/info.php?statistics}},
Facebook has more than $500$ millions users and $65$ billions friendship
relations.

The objective of this paper is to find effective techniques to store and 
access large graphs that can be applied fruitfully not only to web graphs 
but also to social networks of other kinds.
The considerations above explain why this problem is lately emerging as one of 
the central algorithmic issues in the field of information 
retrieval~\cite{citeulike:3908768,CKLCSN}; it should also be noted that improving 
the compression performance on a class of networks, apart for its obvious 
practical consequences, implies (and requires) a better understanding of the 
regularities and of the very structure of such networks.



Here and in the following, we are thinking of \emph{compressed data
structures}. A compressed data structure for a graph must provide very fast amortised 
random access to an edge (link), say in the order of few hundreds of nanoseconds, 
as opposed to a ``compression scheme'', whose only evaluation criterion is the number of bits per link. 
While this definition is not formal, it excludes methods in which the successors of a
node are not accessible unless, for instance, a large part of the graph is scanned.
In a sense, compressed data structures are the empirical counterpart of
\emph{succinct} data structures (introduced by Jacobson~\cite{JacSSTG}), which
store data using a number of bits equal to the information-theoretical lower
bound, providing access asymptotically equivalent to a standard data structure.

The idea of using a compressed data structure to store social networks was
already successfully exploited with application to web graphs~\cite{BoVWFI}, 
showing that such graphs may be stored using less than 3 bits/link; this
impressive compression ratio is mostly obtained by making good use of two simple properties 
that can be experimentally observed when nodes are ordered lexicographically by 
URL~\cite{RSWLD}: 
\begin{itemize}\setlength{\itemsep}{0pt}
  \item \emph{similarity}: nodes that are close to each other in the order tend
  to have similar sets of neighbours;
  \item \emph{locality}: most links are between nodes that are close to each
  other in the order.
\end{itemize}

The fact that most compression algorithms exploit these (or analogous)
properties explains why such algorithms are so sensible to the way nodes are
ordered; the solution of ordering nodes lexicographically by URL
is usually considered good enough for all practical purposes, and has the extra 
advantage that even the URL list can be compressed very efficiently via prefix 
omission. Analogous techniques, which use additional information besides the 
graph itself, are called \emph{extrinsic}. One natural and important question is 
whether there exist any \emph{intrinsic} order of the nodes (i.e., one that does 
not rely on any external data) that produces comparable, or maybe even better, 
compression ratios. This is particularly urgent for general social networks, 
where the very notion of URL does no longer apply and finding a natural
extrinsic order is problematic~\cite{CKLCSN,BSVPWG}.

\section{Problem Definition and Related Works}
\label{sec:PR}
The general problem we consider may be stated as follows: a graph-compression
algorithm $\mathscr A$ takes (the adjacency matrix of) a graph as input and
stores it in a compressed data structure; the algorithm output depends on the specific 
numbering chosen for the nodes. We let $\rho_{\mathscr A}(G,\pi)$ be the number of 
bits per link needed by $\mathscr A$ to store the graph $G$ under the given node 
numbering\footnote{Throughout this paper, we use von Neumann's
notation $n=\{\,0,1,\dots,n-1\,\}$.} $\pi: V_G \to |V_G|$. The overall objective
is to find a numbering $\hat\pi$ minimising $\rho_{\mathscr A}(G,\hat\pi)$. 
In the following, we shall always assume that a graph $G$ with $n$ nodes has
$V_G=n$, so a node numbering is actually a permutation $\pi:n\to n$.

Of course, the problem has different solutions depending on the specific
compression algorithm $\mathscr A$ that is taken into consideration. In the
following, we shall focus on the so-called BV compression scheme~\cite{BoVWFI} used within the
WebGraph framework, which incorporates the main ideas adopted in earlier
systems and is a \textit{de facto} standard for handling large web-like
graphs. In particular, the framework strongly relies on similarity and locality
to achieve its good compression results; for this reason, we believe that most
compressed structures that are based on the same properties will probably
display a similar behaviour.

As noted in~\cite{CKLCSN}, even a very mild version of the
above-stated optimisation problem turns out to be NP-hard, so we can only
expect to devise heuristics that work well in most practical cases.
Such heuristics may be intrinsic or extrinsic, depending on whether they only
use the information contained in the graph itself or they also depend on some
external knowledge.

In the class of intrinsic order heuristics,~\cite{RSWLD} proposes to choose the
permutation $\pi$ that would sort the rows of the adjacency matrix $A_G$ in
lexicographic order. This is an example of a more general kind of solution: fix
some total ordering $\prec$ on the set of $n$-bit vectors (e.g., the
lexicographic ordering), and let $\pi$ be the permutation that would sort the
rows of the adjacency matrix $A_G$ according to\footnote{Here we are
disregarding the problem that $\pi$ is not unique if the adjacency matrix contains
duplicated rows. This issue turns out to have a negligible impact on compression 
and will be ignored in the following.} $\prec$. 

Another possible solution in the same class, already mentioned in~\cite{RSWLD} 
and studied more deeply in~\cite{BSVPWG}, consists in letting $\prec$ be a Gray 
ordering. Recall that~\cite{KnuthTuples} an \emph{$n$-bit Gray ordering} is a total order
on the set of the $2^n$ binary $n$-bit vectors such that any two successive
vectors differ in exactly one position. Although many $n$-bit Gray ordering exist, a very
effective one (i.e., one that is manageable in practice because it is easy to
decide which of two vectors come first in the order) is the so-called
\emph{reflective $n$-bit Gray ordering}, which was used
in~\cite{BSVPWG}.\footnote{Since in the rest of this paper we will only deal with
this Gray ordering, we will simply omit the adjective ``reflective'' in the
following.}

Chierichetti \textit{et al.}~\cite{CKLCSN} propose a completely different
intrinsic approach based on shingles that adopts ideas used for document similarity
derived from min-wise independence. The compression results they get are
comparable to those achieved through Gray ordering~\cite{BSVPWG}. 
In the same paper they also discuss an alternative compression technique
(called BL) that provides better ratios; however, while interesting as a compression
scheme, BL does not provide a compressed data structure---recovering the
successors of a node requires, in principle, decompressing the whole graph.

Recently, Safro and Temkin~\cite{SaTMANCO} presented a multiscale approach 
for the network minimum logarithmic arrangement problem: their method searches
for an intrinsic ordering that optimises directly the sum of the logarithms 
of the gaps (numerical difference between two successive neighbours). Although
their work is not aimed at compression, their ordering is potentially useful
for this task if combined with a compression scheme like BV. Indeed, some
preliminary tests show that these orderings are promising especially on social
networks; however, their implementation does not scale well to datasets
with more that a few millions of nodes and so it is impractical for our purpose.

As far as extrinsic orderings are concerned, a central r\^ole is played by the
URL-based ordering in a web graph. If $G$ is a web graph, we can assume to have a permutation $\pi_U$ of its nodes
that sorts them according to the lexicographic URL ordering: this extrinsic
heuristic dates back to~\cite{BBHCS} and, as explained above, turns out to give
very good compression, but it is clearly of no use in non-web social networks.
Another effective way to exploit the host information is presented in~\cite{BSVPWG}, 
where URLs from the same host are kept adjacent 
(within the same host, Gray ordering is used instead).

It is worth remarking that all the intrinsic techniques mentioned above produce
different results (and, in particular, attain different compression ratios) depending on the
\emph{initial} numbering of the nodes, because they work on the adjacency matrix
$A_G$. This fact was overlooked in almost all previous literature, but it
turns out to be very relevant: applying one of these intrinsic re-ordering to a randomly 
numbered graph (see Table~\ref{tab:bits_random}) produces worse compression
ratios than starting from a URL-ordered web graph (see 
Table~\ref{tab:bits_nonrandom}). 

This problem arises because even if the intrinsic techniques described above
do not explicitly use any external information, the initial order of a
graph is often obtained by means of some external information, so the
compression performances cannot be really considered intrinsic. To make this point clear, 
throughout the paper we will speak of \emph{coordinate-free} algorithms for those 
algorithms that achieve almost the same compression performances starting from any 
initial ordering; this adjective can be applied both to compression algorithms and 
to orderings+compression algorithm pairs. From an experimental viewpoint, 
this means that, unlike in the previous literature, we run all our
tests starting from a \emph{random permutation} of the original graph. We
suggest this approach as a baseline for future research, as it avoids 
any dependency on the way in which the graph is presented initially.

The only coordinate-free compression algorithm we are aware of\footnote{The
quite extensive survey in~\cite{BSVPWSG} shows that many other approaches to
web-graph compression, not quoted here, either fail to compress social networks,
or are strongly dependent on the initial ordering of the graph.} is that
proposed by Apostolico and Drovandi in~\cite{ApDGCB};\footnote{Our experiments
show in fact a very limited variation in compression (10--15\%) when starting
from URL ordering or from a random permutation, except for the
\texttt{altavista-nd} dataset, which however is quite pathological.} they
exploit a breadth-first search (BFS) to obtain an ordering of the graph and they
devise a new compression scheme that takes full advantage of it. Their algorithm
has a parameter, the \emph{level}, which can be tuned to obtain different
trade-offs between compression performance and time to retrieve the adjacency
list of a node: at level 8 they attain better compression performances than
those obtained by BV with Gray orderings and have a similar speed in retrieving
the adjacency list. Even in this optimal setting, though, their approach is
outperformed by the one we are going to present (see Table~\ref{tab:llpavsad}).

Finally, Maserrat and Pei~\cite{MPNQFC} propose a completely different approach
that does not rely on a specific permutation of the graph. Their method
compresses social networks by exploiting Eulerian data structures and
multi-position linearisations of directed graphs. Notably, their technique is
able to answer both successor and predecessor queries: however, while querying for
adjacency of two nodes is a fast operation, the cost per link of enumerating 
the successors and predecessors of a node is between one and two orders of
magnitude larger than what we allowed. In other words, by the
standards followed in this paper their algorithm does not qualify as a
compressed data structure.

We must also remark that the comparison given in~\cite{MPNQFC} of the
compression ratio w.r.t.~WebGraph's BV scheme is quite unfair: indeed, the
authors argue that since their algorithm provides both predecessors and
successors, the right comparison with the BV scheme requires roughly doubling
the number of bits per link (as the BV scheme just returns successors). However,
this bound is quite na\"ive: consider  a simple strategy that uses the set
$E_{\text{sym}}$ of all symmetric edges, and let $G_{\text{sym}} =
(V,E_{\text{sym}})$ and $G_{\text{res}} = (V,E \setminus E_{\text{sym}})$. To be
able to answer both successor and predecessor queries one can just store
$G_{\text{sym}}$, $G_{\text{res}}$ and $G_{\text{res}}$ transposed. Using this
simple strategy and applying the ordering proposed in this paper to the datasets
used in~\cite{CKLCSN} we obtain better compression ratios.


\section{Our Contribution}

In this paper we give a number of algorithmic and experimental results:
\begin{itemize}
  \item We identify two measures of fitness for algorithms that try to recover
  the host structure of the web, and report experiments on large web graphs
  that suggest that the success of the best coordinate-free
  orderings is probably due to their capability of guessing the host structure.
  \item Since the existing coordinate-free orderings do not work well on social
  networks, we propose a new algorithm, called \emph{Layered Label
  Propagation}, that builds on previous work on scalable
  clustering by label
  propagation~\cite{citeulike:1724653,RoNLRPMCD}; the algorithm
  can reorder very large graphs (billions of nodes), and unlike
  previous proposals, is free from parameters.
  \item We report experiments on the compression of a wide array of web graphs
  and social networks using WebGraph after a reordering by Layered Label
  Propagation; the experiments show that our combination of techniques provides
  a major increase in compression with respect to all currently known
  approaches. This is particularly surprising in view of the fact that we obtain
  the best results \emph{both} on web graphs \emph{and} on social networks.
  Our largest graph contains more than 600 millions nodes---one order of magnitude
  more than any published result in this area.
\end{itemize}

Almost all the datasets can be downloaded from
the site \texttt{http://law.dsi.unimi.it/} (or from other public
or free sources) and have been widely used in the previous literature to benchmark
compression algorithms. The Java code for our new algorithm is distributed
at the same URL under the GNU General Public License.

We remark that our new algorithm has also been applied with excellent results to
the Minimum Logarithmic Arrangement
Problem~\cite{SaTMANCO}~\footnote{\texttt{http://www.mcs.anl.gov/\textasciitilde
safro/mloga.html}. The authors had been provided a preliminary version of our code to perform their tests.}.

\section{Recovering Host information \\from a Random Permutation}

As a warm-up towards our new algorithm, we propose an empirical analysis that
aims at determining objectively why existing approaches compress well web
graphs.

The results presented in~\cite{BSVPWG} suggest that what is really 
important in order to achieve good compression performances on web graphs is not 
the URL ordering \emph{per se}, but rather an ordering that keeps nodes from 
the same host close to one another. For this reason, we will be naturally
interested to measure how much a given ordering $\pi$ respects the partition
induced by the hosts, $\mathscr{H}$.

The \emph{first measure} we propose is the probability to have a \emph{host
transition} (HT): 
\[
\operatorname{HT}(\mathscr{H},\pi) =  \frac{ \sum_{ i = 1 }^{|V_G| - 1}
\delta\left(
\mathscr{H}[ \pi^{-1} (i) ] , \mathscr{H}[ \pi^{-1} (i-1) ] \right) }{|V_G|-1}
\]
where  $\delta$ denotes the usual Kronecker's delta
and $\mathscr{H}[x]$ is the equivalence class of node $x$ (i.e., the set of all
nodes that have the same host as $x$): this is simply the fraction of
nodes that are followed, in the order $\pi$, by another node with a different host.

Alternatively, we can reason as follows: the ordering induces a refinement of
the original host partition, and the appropriateness of a given ordering can be
measured by comparing the original partition with the refined one. 
More formally, let us denote with $\mathscr{H}_{|\pi}$ the partition induced by the
reflexive and transitive closure of the relation $\rho$ defined by
$$ x \;\rho\; y \iff |\pi(x)-\pi(y)| = 1 \text{ and } \mathscr{H}[x] =
\mathscr{H}[y].$$ Intuitively, the classes of $\mathscr{H}_{|\pi}$ are made of
nodes belonging to the same host and that are separated in the order only by nodes of the same host. 
Notice that this is always a refinement of the partition $\mathscr{H}$. 

The \emph{second measure} that we have decided to employ to compare partitions
is the Variation of Information (VI) proposed in~\cite{MeiCCAV}. Define the entropy
associated with the partition $\mathscr{S}$ as:
\[
H(\mathscr{S}) = - \sum_{S\in\mathscr S} P(S) \log(P(S)) \quad \text{where} \; P(S) = \frac{|S|}{|V_G|}
\]
and the mutual information between two partitions as:
\[
I(\mathscr{S},\mathscr{T}) = \sum_{S\in\mathscr S} \sum_{T\in\mathscr T} P(S,T) \log \frac{P(S,T)}{P(S)P(T)}  
\]
where $P(S,T) = \frac{|S\cap T|}{|V_G|}$. The Variation of information is then
defined as
\[
VI(\mathscr{S},\mathscr{T}) = H(\mathscr{S}) + H(\mathscr{T}) - 2\, I(\mathscr{S},\mathscr{T});
\]
notice that, in our setting, since $\mathscr{H}_{|\pi}$ is always a refinement
of $\mathscr{H}$, we have $I(\mathscr{H},\mathscr{H}_{|\pi}) = H(\mathscr{H})$
and so VI simplifies into
\[
VI(\mathscr{H},\mathscr{H}_{|\pi}) = H(\mathscr{H}_{|\pi}) - H(\mathscr{H}). 
\]

\medskip
Armed with these definitions, we can determine how much different intrinsic
orderings are able to identify the original host structure. We computed the two
measures defined above on a number of web graphs (see Section~\ref{sec:exp}) and
using some different orderings described in the literature; more precisely, we
considered:
\begin{itemize}
  \item \emph{Random}: a random node order;
  \item \emph{Natural}: for web graphs, this is the URL-based ordering; for the
  other non-web social networks, it is the order in which nodes are presented,
  which is essentially arbitrary (and indeed produces compression ratios not
  very different from random);
  \item \emph{Gray}: the Gray order explained in~\cite{BSVPWG};
  \item \emph{Shingle}: the compression-friendly order described in~\cite{CKLCSN};
  \item \emph{BFS}: the breadth-first search traversal order, exploited
  in~\cite{ApDGCB};
  \item \emph{LLP}: the Layered Label Propagation algorithm described in
  this paper (see Section~\ref{sec:llpa} for details).
\end{itemize}

\begin{table*}[t]
	\begin{center}
		
\begin{tabular}{ l cccccccccccc }
\toprule
\textbf{Name} & \multicolumn{2}{c}{\textbf{LLP}}&\multicolumn{2}{c}{\textbf{BFS}}&\multicolumn{2}{c}{\textbf{Shingle}}&\multicolumn{2}{c}{\textbf{Gray}}&\multicolumn{2}{c}{\textbf{Natural}}&\multicolumn{2}{c}{\textbf{Random}} \\
& \textbf{HT}&\textbf{VI}&\textbf{HT}&\textbf{VI}&\textbf{HT}&\textbf{VI}&\textbf{HT}&\textbf{VI}&\textbf{HT}&\textbf{VI}&\textbf{HT}&\textbf{VI} \\ \midrule
\texttt{eu} & 1.58\%&4.60&2.04\%&4.60&20.12\%&7.33&20.09\%&7.55&0.05\%&0.00&97.11\%&13.80 \\
\texttt{in} & 1.83\%&1.92&2.53\%&2.32&15.83\%&4.51&37.11\%&6.76&0.32\%&0.00&99.62\%&11.37 \\
\texttt{indochina} & 1.37\%&1.61&1.99\%&2.63&32.05\%&6.03&30.96\%&5.93&0.26\%&0.00&99.93\%&11.71 \\
\texttt{it} & 3.05\%&2.63&2.93\%&2.83&27.04\%&5.32&26.18\%&5.27&0.34\%&0.00&99.99\%&11.45 \\
\texttt{uk} & 2.52\%&2.88&1.29\%&2.65&20.64\%&5.52&19.93\%&5.46&0.11\%&0.00&99.98\%&13.76 \\
\bottomrule
\end{tabular}

	\end{center}
	\caption{Various measures to evaluate the ability of different orderings to
	recover host information. Smaller values indicate a better recovery.}
	\label{tab:host_meas}
\end{table*}

The results of this experiment are shown in Table~\ref{tab:host_meas}; comparing
them with the compression results of Table~\ref{tab:bits_random} (that shows the
compression performances starting from a truly random order), it is clear that
recovering the host structure from random is the key property that is needed for
obtaining a real coordinate-free algorithm. However, the only ordering proposed
so far that is able to do this is breadth-first search, and its capability to
identify hosts seems actually a side effect of the very structure of the web. In
the rest of the paper, we use BFS as a strong baseline against which our new
results should be compared.\footnote{It is unlikely that, in presence of the
more complicated structure that we expect in social networks, an algorithm as simple as a breadth-first search can identify meaningful clusters (see again the BFS
column of Table~\ref{tab:bits_random}), and this leaves room for improvement.}.
 
\section{Label Propagation Algorithms}
\label{sec:lpa}

Most of the intrinsic orderings proposed so far in the literature are unable to
produce satisfactory compression ratios when applied to a randomly permuted
graph, mainly because they mostly fail in reconstructing host information as we
discussed in the last section. To overcome their limitations, we can try to
approach this issue as a clustering problem. However, this attempt presents a
number of difficulties that are rather peculiar. First of all, the size of the
graphs we are dealing with imposes to use algorithms that scale linearly with
the number of arcs (and there are very few of them; see~\cite{ForCDG}). Moreover,
we do not possess any prior information on the number of clusters we should expect and their sizes are going to be highly unbalanced.

These difficulties strongly restrict the choice of the clustering algorithm. 
In the last years, a new family of clustering algorithms were 
developed starting from the label propagation algorithm presented 
in~\cite{citeulike:1724653}, that use the network structure alone as their guide 
and require neither optimisation of a predefined objective 
function nor prior information about the communities. These algorithms are
inherently local, linear in the number of edges, and require just few passes
on the graph.

The main idea of label propagation algorithms is the following: the algorithms
execute in rounds, and at the beginning of each round every node has a label
representing the cluster that the node currently belongs (at the beginning,
every node has a different label).
At each round, every node will update its label according to some rule, the
update order being chosen at random at the beginning of the round; the algorithm
terminates as soon as no more updates take place. Label propagation algorithms
differ from each other on the basis of the update rule.

The algorithm described in~\cite{citeulike:1724653} (hereafter referred to as
\emph{standard label propagation} or just \emph{label propagation}) works
on a purely local basis: every node takes the label that occurs more frequently in its neighbourhood\footnote{In the case of ties, a random choice is performed,
unless the current label of the node is one of the most frequent in its
neighbourhood, in which case the label is simply not changed.}. Metaphorically,
every node in the network chooses to join the largest neighbouring community
(i.e., the one to which the maximum number of its neighbours belongs). As labels
propagate, densely connected groups of nodes quickly reach a consensus on 
a unique label. When many such dense consensus groups are created throughout 
the network, they continue to expand outwards until it is possible to do so. 
At the end of the propagation process, nodes having the same 
labels are grouped together as one community.

It has been proved~\cite{2008PhyA..387.4982T} that this kind of
label propagation algorithm is formally equivalent to finding the local minima of the 
Hamiltonian for a kinetic Potts model. This problem has a trivial globally
optimal solution when all the nodes have the same label; nonetheless, since 
the label-propagation optimisation procedure
produces only local changes, the search for maxima in the Hamiltonian is prone 
to becoming trapped at a local optimum instead of reaching the global optimum. 
While normally a drawback of local search algorithms, this characteristic 
is essential to clustering: the trivial optimal solution is avoided by the 
dynamics of the local search algorithm, rather than through formal exclusion.

Despite its efficiency, it was observed 
that the algorithm just described tends to produce one giant cluster containing
the majority of nodes. 
The presence of this giant component is due to the very topology of social 
networks;
to try to overcome this problem we have tested variants of the label propagation 
that introduce further constraints. One of the most interesting is the algorithm 
developed in~\cite{PhysRevE.80.026129}, where the update rule is modified in 
such a way that the objective function being optimised becomes the 
\emph{modularity}~\cite{PhysRevE.69.026113} of the resulting clustering. 
Unfortunately, modularity is not a good measure in very large graphs
as pointed out by several authors (e.g., \cite{2007PNAS..104...36F}) due to its
resolution limit that makes it hardly usable on large networks.

Another variant, called Absolute Pott Model (APM)~\cite{RoNLRPMCD},
introduces a nonlocal discount based on a resolution parameter
$\gamma$. For a given node $x$, let
$\lambda_1,\dots,\lambda_k$ be the labels currently appearing on the neighbours of $x$, $k_i$ be the number of neighbours
of $x$ having label $\lambda_i$ and $v_i$ be the overall number of nodes in the
graph with label $\lambda_i$; when $x$ is updated, instead of choosing the label
$\lambda_i$ maximizing $k_i$ (as we would do in standard label propagation),
we choose it as to maximise (see Algorithm~\ref{algo:lpa}) \[ k_i-\gamma(v_i-k_i). \] Observe
that when $\gamma=0$ the algorithm degenerates to label propagation; the
reason behind the discount term is that when we decide to join a given community, we are increasing
its density because of the $k_i$ new edges joining $x$ to existing members of the
community, but we are at the same time decreasing it because of $v_i-k_i$
non-existing edges. Indeed, it can be shown that the density of the sparsest
community at the end of the algorithm is never below $\gamma/(\gamma+1)$.

\begin{algorithm}
\caption{\label{algo:lpa}The APM algorithm.
$\lambda$ is a function that will provide, at the end, the cluster labels. For the sake of
readability, we omitted the resolution of ties.} 
\begin{algorithmic}[1] 
\REQUIRE $G$ a graph, $\gamma$ a density parameter
\STATE $\pi \leftarrow$ a random permutation of $G$'s nodes  
\STATE for all $x$: $\lambda(x)\leftarrow x$, $v(x)\leftarrow 1$
\WHILE{(some stopping criterion)}
	\FOR{$i=0,1,\ldots,n-1$}
		 \STATE for every label $\ell$, $k_\ell\leftarrow
		|\lambda^{-1}(\ell)\cap N_G(\pi(i))|$\\
        \STATE $\hat\ell\leftarrow  \operatorname{argmax}_\ell
        [k_\ell-\gamma(v(\ell)-k_\ell)]$\\ 
        \STATE decrement $v(\lambda(\pi(i)))$
        \STATE $\lambda(\pi(i))\leftarrow \hat\ell$
        \STATE increment $v(\lambda(\pi(i)))$
    \ENDFOR
\ENDWHILE
\end{algorithmic}
\end{algorithm}

\begin{figure}[htb]
\includegraphics[scale=.3]{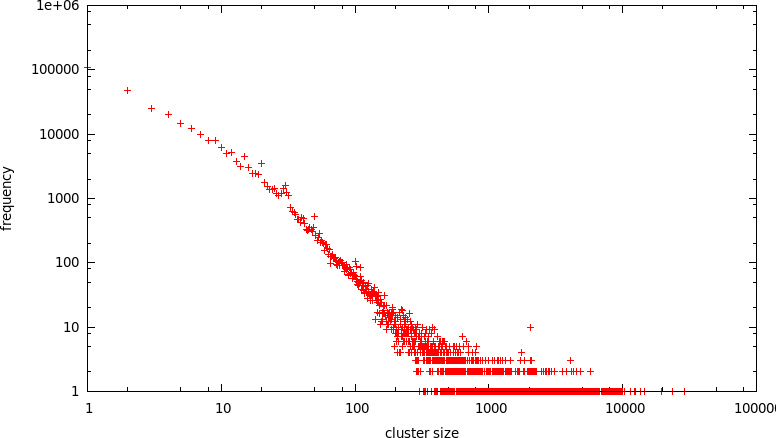}
\caption{\label{fig:heavy}An example of the distribution of cluster sizes
computed by APM.}
\end{figure}

This algorithm demonstrated to be the best candidate for our needs. However it 
has two major drawbacks. The first is that there are no theoretical results 
that can be used to determine \emph{a priori} the optimal value of
$\gamma$ (on the contrary, experiments show that such an optimal value is
extremely changeable and does not depend on some obvious parameters like the
network size or density). The second is that it tends to produce clusters with 
sizes that follow a heavy-tailed decreasing distribution, 
yielding both a huge number of clusters and clusters with a huge number of nodes
(see Figure~\ref{fig:heavy}). Thus to obtain good compression performances we
have to decide both the order between clusters and the order of the nodes that belong to the same cluster.

\section{Layered Label Propagation}
\label{sec:llpa}


In this section we present a new algorithm based on label propagation
that yields a compression-friendly ordering. 

A run of the APM
algorithm (discussed in the previous section) over a given graph and with a
given value of the parameter $\gamma$ produces as output a clustering, that
may be represented as a labelling (mapping each node to the label of the cluster it belongs to). 
An important observation is that, intuitively, there is no notion of optimality
for the tuning of $\gamma$: every value of this parameter describes a different 
resolution of the given graph. 
Values of $\gamma$ close to $0$ highlight a coarse structure with few, big
and sparse clusters, while, as $\gamma$ grows, the clusters are small and
dense, unveiling a fine-grained structure. Ideally, we would like to find a way to
compose clusterings obtained at different resolution levels.\footnote{Of
course, such a compositional approach could be applied also to other scalable
clustering techniques: we have experimented with several
alternatives~\cite{ForCDG}, and APM is by far the most interesting candidate.}

This intuition leads to the definition of \emph{Layered Label
Propagation (LLP)}; this algorithm is iterative and produces a sequence of node
orderings; at each iteration, the APM algorithm is run with a suitable value of $\gamma$ and the
resulting labelling is then turned into an
ordering of the graph that keeps nodes with the same label close to one another;
nodes within the \emph{same cluster} are left in the same order they had before. 

To determine the relative order among \emph{different clusters}, it is worth
observing that the \emph{actual label} produced by the label propagation algorithm suggests a
natural choice: since every cluster will be characterised by the initial label
of the leader node (the node which flooded that portion of graph; see
Algorithm~\ref{algo:lpa}), we can sort the clusters according to the order that
the leader nodes had.

More formally, let a sequence $\gamma_0,\gamma_1,\gamma_2,\ldots$ and an initial
ordering $\pi_0: V_G \to |V_G|$ of the nodes of $G$ be fixed; 
we define a sequence of orderings $\pi_1,\pi_2,\ldots: V_G
\to |V_G|$ and a sequence of labelling functions $\lambda_0,\lambda_1,\ldots:
V_G \to |V_G|$ as follows: $\lambda_k$ is obtained by running the APM algorithm on the graph $G$
with parameter $\gamma_k$; then we let $\pi_{k+1}$ be the ordering
defined by
\[
 x \leq_{k+1} y \text{ iff }
	\begin{cases} 
		\pi_k(\lambda_k(x)) < \pi_k(\lambda_k(y)) \quad \text{or}\\
	  	\lambda_k(x) = \lambda_k(y) \; \bigwedge \; \pi_k(x) \leq
	  	\pi_k(y).
	\end{cases}
\]

The rationale behind this way of composing the newly obtained clustering with
the previous ordering is explained above: elements in the same cluster (i.e.,
with the same label) are ordered as before; for elements
with a \emph{different} label, we use the order that the corresponding labels
(i.e., leader nodes) had before.

The output of LLP actually depends on two elements: the initial
ordering $\pi_0$ and the choice of the parameters $\gamma_k$ at each iteration.

Regarding the choice of the $\gamma_k$'s, instead of trying to find at each
iteration an optimal value for the parameter we exploit the diverse resolution
obtained through different choices of the parameter, thus finding a proper order
between clusters that suitably mixes the clusterings obtained at all resolution levels.
To obtain this effect, we choose every $\gamma_k$ uniformly at random in
the set\footnote{Although in theory $\gamma$ could be larger than 1, such a
choice would be of no practical use on large networks, because it would
only yield a complete fragmentation of the graph.} $\{0\}\cup\{2^{-i},
i=0,\dots,K\}$. Since the APM algorithm is run at
every step on the same graph $G$, it turns out that it is easier (and more efficient) 
to precompute the labelling function output by the APM algorithm for each $\gamma$ in the above set, and then to
re-use such labellings. 

The surprising result is that the final ordering obtained by this mutilresolution
strategy is better than the ordering obtained by applying the same strategy with 
$K$ different clusterings generated with the same value of $\gamma$ chosen after 
a grid search for the optimal value (as shown in
Table~\ref{tab:llpavsfixedllpa}), and \emph{a fortiori} on the ordering induced
by one single clustering generated with the optimal $\gamma$. 
Moreover the final order obtained is essentially independent on the
initial permutation $\pi_0$ of the graph (as one can see comparing Table~\ref{tab:bits_nonrandom}
with Table~\ref{tab:bits_random}).

One may wonder if this iterative strategy can be 
applied also to improve the performances of other intrinsic orderings. Our
experiments rule out this hypothesis. Iterating Gray, lex, or BFS orderings
does not produce a significant improvement. 

\begin{table}[t]
	\begin{center}
		
\begin{tabular}{ l r r r r }
\toprule
\textbf{Name} &\multicolumn{2}{c}{\textbf{LLP}} & \multicolumn{2}{c}{\textbf{Fixed LLP}} \\ \midrule
\texttt{Amazon} & 9.12 & & 9.43 & (+3\%) \\
\texttt{DBLP} & 6.87 & & 7.13 & (+3\%) \\
\texttt{Enron} & 6.45 & & 6.90 & (+6\%) \\
\texttt{Hollywood} & 5.17 & & 5.55 & (+7\%) \\
\texttt{LiveJournal} & 10.95 & & 11.40 & (+4\%) \\
\texttt{Flickr} & 8.9 & & 9.27 & (+4\%) \\
\texttt{indochina (hosts)} & 5.57 & & 6.25 & (+12\%) \\
\texttt{uk (hosts)} & 6.35 & & 6.79 & (+6\%) \\
\midrule
\texttt{eu} & 3.88 & & 4.46 & (+14\%) \\
\texttt{in} & 2.44 & & 2.99 & (+22\%) \\
\texttt{indochina} & 1.68 & & 1.92 & (+14\%) \\
\texttt{it} & 2.05 & & 2.59 & (+26\%) \\
\texttt{uk} & 1.8 & & 2.27 & (+26\%) \\
\bottomrule
\end{tabular}

	\end{center}
	\caption{Comparison between LLP with different values of
	$\gamma$ and LLP with the best value of
	$\gamma$ only. Values are bits per link.}
	\label{tab:llpavsfixedllpa}
\end{table}

\section{Parallel Implementation}

Layered label propagation lends itself naturally to the
\emph{task-de\-com\-position} parallel-programming paradigm, which may dramatically improve performances on
modern multicore architectures: since the update order is randomised, there is
no obstacle in updating several nodes in parallel. Our implementation breaks the
set of nodes into a very small number of tasks (in the order of thousands). A large number of threads picks up the first available task and solves it: as
a result, we obtain a performance improvement that is linear in the number of
cores. We are helped by WebGraph's facilities, which allows us to provide 
each thread with a lightweight copy of the graph that shares the bitstream and
associated information with all other threads.

\section{Experiments}
\label{sec:exp}
For our experiments, we considered a number of graphs with various sizes and
characteristics; most of them are (directed or undirected) social graphs of
some kind, but we also considered some web graphs for comparison (because for
web graphs we can rely on the URLs as external source of information). More
precisely, we used the following datasets (see also Table~\ref{tab:social}
and~\ref{tab:web}):
\begin{itemize}\setlength{\itemsep}{0pt}
\item \emph{Hollywood}: 
One of the most popular \emph{undirected} social graphs, the graph of
movie actors: vertices are actors, and two actors are joined
by an edge whenever they appeared in a movie together. 
\item \emph{DBLP}:
DBLP\footnote{\url{http://www.informatik.uni-trier.de/~ley/db/}} is a bibliography 
service from which an \emph{undirected} scientific collaboration network can be
extracted: each vertex of this undirected graph represents a scientist and two
vertices are connected if they have worked together on an article.   
\item \emph{LiveJournal}:
LiveJournal\footnote{\url{http://www.livejournal.com/}} is a
virtual community social site started in 1999: nodes are
users and there is an arc from $x$ to $y$ if $x$ registered $y$ among his
friends (it is not necessary to ask $y$ permission, so the graph is
\emph{directed}). We considered the same 2008 snapshot of \emph{LiveJournal}
used in~\cite{CKLCSN} for their experiments\footnote{The dataset was kindly
provided by the authors of~\cite{CKLCSN}.}.
\item \emph{Amazon}: This dataset describes similarity among books as reported
by the Amazon store; more precisely the data was
obtained\footnote{\url{http://www.archive.org/details/amazon_similarity_isbn/}}
in 2008 using the Amazon E-Commerce Service APIs using
\texttt{SimilarityLookup} queries.
\item \emph{Enron}:
This dataset was made public by the Federal Energy Regulatory Commission during
its investigations: it is a partially anonymised corpus of e-mail messages
exchanged by some Enron employees (mostly part of the senior management). We
turned this dataset into a \emph{directed} graph, whose nodes represent people
and with an arc from $x$ to $y$ whenever $y$ was the recipient of (at least) a
message sent by $x$.
\item \emph{Flickr}:
Flickr\footnote{\url{http://www.flickr.com/}; we thank Yahoo!\ for the
experimental results on the Flickr graph.} is an online community where users
can share photographs and videos. In Flickr the notion of acquaintance is modelled through
\emph{contacts}; we used an \emph{undirected} version of this network, where
vertices correspond to users and there is an edge connecting $x$ and $y$
whenever either vertex is recorded as a contact of the other one.
\item For comparison, we considered five web graphs of various sizes (ranging
from about $800$ thousand nodes to more than $650$ million nodes), available at
the LAW web site  \url{http://law.dsi.unimi.it/}.
\item Finally, the \emph{altavista-nd} graph was obtained from the
Altavista dataset distributed by Yahoo! within the Webscope 
program (AltaVista webpage connectivity dataset, version 1.0\footnote{ 
\url{http://research.yahoo.com/Academic_Relations}}). With respect to the
original dataset, we pruned all dangling nodes (``nd'' stands for ``no
dangling''). The original graph, indeed, contains $53.74$\% dangling nodes (a
preposterous percentage~\cite{VigSMCH}), probably because it also considers the
\emph{frontier} of the crawl---the nodes that have been discovered but not
visited. We eliminated (one level of) dangling nodes to approximate the set of
visited nodes, and also because dangling nodes are of little importance in
compression.\footnote{It should be
remarked by this graph, albeit widely used in the literature, is not a good
dataset. As we already noted, most likely all nodes in the
frontier of the crawler (and not only visited nodes) were added to the graph;
moreover, the giant component is less than 4\% of the whole graph.}

\end{itemize}

\begin{table}[t]
	\begin{center}
		
\begin{tabular}{ l r r }
\toprule
\textbf{Name} & \textbf{Nodes} & \textbf{Edges} \\ \midrule
\texttt{Amazon} & 735\,323 & 5\,158\,388 \\
\texttt{DBLP} & 326\,186 & 1\,615\,400 \\
\texttt{Enron} & 69\,244 & 276\,143 \\
\texttt{Hollywood} & 1\,139\,905 & 113\,891\,327 \\
\texttt{LiveJournal} & 5\,363\,260 & 79\,023\,142 \\
\texttt{Flickr} & 526\,606 & 47\,097\,454 \\
\bottomrule
\end{tabular}

	\end{center}
	\caption{Social graph description.}
	\label{tab:social}
\end{table}

\begin{table}[h]
	\begin{center}
		
\begin{tabular}{ l r r r }
\toprule
\textbf{Name} & \textbf{Year} & \textbf{Nodes} & \textbf{Edges} \\ \midrule
\texttt{eu} & 2005 & 862\,664 & 19\,235\,140 \\
\texttt{in} & 2004 & 1\,382\,908 & 16\,917\,053 \\
\texttt{indochina} & 2004 & 7\,414\,866 & 194\,109\,311 \\
\texttt{indochina (hosts)} & 2004 & 19\,123 & 233\,380 \\
\texttt{it} & 2004 & 41\,291\,594 & 1\,150\,725\,436 \\
\texttt{uk (hosts)} & 2005 & 587\,205 & 12\,825\,465 \\
\texttt{uk} & 2007 & 105\,896\,555 & 3\,738\,733\,648 \\
\texttt{altavista-nd} & 2002 & 653\,912\,338  &  4\,226\,882\,364 \\
\bottomrule
\end{tabular}

	\end{center}
	\caption{Web graph description.}
	\label{tab:web}
\end{table}

Each graph was compressed in the BV format using
WebGraph~\cite{BoVWFI}\footnote{We adopted
the default window size ($W=7$), disabled intervalisation and put a limit of 3 to the length
of the possible reference chains (see~\cite{BSVPWG} for details on the r\^ole
of this parameter). Observe that the latter two settings tend to deteriorate the
compression results, but make decompression extremely efficient even when
random access is required.} and we measured the compression performance using 
the number of bits/link actually occupied by the graph file. 

We also compared LLP+BV with the compression obtained using the
algorithm proposed by Apostolico and Drovandi~\cite{ApDGCB} at level 8 starting
from a randomly permuted graph; the results, shown in Table~\ref{tab:llpavsad},
provide evidence that LLP+BV outperforms AD in all cases, and in a
significant way on social networks and large web graphs. This is particularly
relevant, since the compression algorithm of AD is designed to take full advantage of a specific ordering (the breadth-first search) and is the only known coordinate-free alternative we are aware of. In our comparison, contrarily
to all other tables, we used the full compression power of the BV format, as our
intent is to motivate LLP+BV as a very competitive coordinate-free
compression algorithm. In the rest of the paper, as we already explained, we
have turned off intervalisation, as our purpose is to study the effect of
different permutations on locality and similarity: this explains why the bits
per link found in Table~\ref{tab:llpavsad} are smaller than elsewhere in the
paper.

A comment is needed about the bad performance the Apostolico--Drovandi
method on the \texttt{altavista-nd} dataset. Apparently, the size of the dataset is
such that scrambling it by a random permutation causes the method to use a bad
naming for the nodes, in spite of the initial breadth-first visit. In our
previous experiments, the Apostolico--Drovandi method did not show variations of
more than 20\% in compression due to random permutations, but clearly the issue
needs to be investigated again.

\begin{table}[t]
	\begin{center}
		\begin{tabular}{ l r r r r }
\toprule
\textbf{Name} &\multicolumn{2}{c}{\textbf{LLP+BV}} & \multicolumn{2}{c}{\textbf{AD}} \\ \midrule
\texttt{Amazon} 		& 9.13 	& & 12.39 	&(+36\%) \\
\texttt{DBLP} 			& 6.82 	& & 7.47 	&(+10\%) \\
\texttt{Enron} 			& 6.07 	& & 7.74 	&(+28\%) \\
\texttt{Hollywood} 		& 4.99 	& & 7.64 	&(+53\%) \\
\texttt{LiveJournal} 		& 10.91 & & 14.97 	&(+37\%) \\
\texttt{Flickr} 		& 8.9 	& & 11.19 	&(+26\%) \\
\texttt{indochina (hosts)} 	& 5.42 	& & 6.83 	&(+26\%) \\
\texttt{uk (hosts)} 		& 6.19 	& & 7.85 	&(+27\%) \\
\midrule
\texttt{eu} 			& 3.78	& & 4.01 	&(+6\%) \\
\texttt{in} 			& 2.24 	& & 2.39 	&(+7\%) \\
\texttt{indochina} 		& 1.53 	& & 1.70 	&(+11\%) \\
\texttt{it} 			& 1.91 	& & 2.31 	&(+21\%) \\
\texttt{uk} 			& 1.72 	& & 2.32 	&(+36\%) \\
\texttt{altavista-nd}		& 5.16	& & 11.04	&(+114\%) \\
\bottomrule
\end{tabular}

	\end{center}
	\caption{Comparison between LLP+BV compression (for this
	particular table, the full set of compression tecniques available in WebGraph
	has been used, including intervalisation) and the algorithm proposed by
	Apostolico and Drovandi (AD) at level 8. Values are bits per link.}
	\label{tab:llpavsad}
\end{table}

\section{Results}
\label{sec:res}

Tables~\ref{tab:bits_nonrandom} and~\ref{tab:bits_random} present the number of
bits per link required by our datasets under the different orderings discussed
above and produced starting from the natural order and
from a random order (the percentages shown in parenthesis give the gain
w.r.t.~breadth-first search ordering).
Here are some observations that the experimental results suggest:
\begin{itemize}
  \item LLP provides always the best compression, with an average gain 
  of $25\%$  with respect to BFS, and largely outperforms both 
  simple Gray~\cite{BSVPWG} and shingle orderings~\cite{CKLCSN}.
  Some simple experiments not reported
  here shows that the same happen for transposed graphs: for instance,
  \texttt{uk} is compressed at $1.06$ bits per link. This makes LLP+BV
  encoding by far the best compressed data structure available today.
  \item LLP is extremely robust with respect to the initial ordering of
  nodes and its combination with BV provides actually a coordinate-free
  compressed data structure. Other orderings (in particular, Gray and shingle) are much
  more sensitive to the initial numbering, especially on web graphs. We urge researchers in this
  field \emph{to always generate permutations starting from a randomised copy of the graph}, 
  as ``useful'' ordering information in the original dataset can percolate as
  an artifact in the final results. 
  
  \item As already remarked elsewhere~\cite{CKLCSN}, social networks seem
  to be harder to compress than web graphs: this fact would suggest that
  there should be some yet unexplained topological difference between the two kinds of
  graphs that accounts for the different compression ratio.
\end{itemize}

Despite the great improvement in terms of compression results our technique
remains highly scalable. All experiments are performed on a Linux server
equipped with Intel Xeon X5660 CPUs ($2.80$\,GHz, $12$\,MB cache size) for
overall 24 cores and $128$\,GB of RAM; the server cost about $8\,900$ EUR in
2010. Our Java implementation of LLP sports a linear scaling in the
number of arcs with an average speed of $\approx 80\,000\,000$\,arcs/s per
iteration. The overall time cost of the algorithm depends on the number
$\gamma$'s and on the stopping criterion. With our typical setting the overall
speed of the algorithm is $\approx 800\,000$\,arcs/s.
  
The algorithm is also very memory efficient (it uses $3n$ integers plus the
space required by the graph\footnote{It is possible in principle to avoid
keeping the graph in main memory, but the cost becomes $O(n\log n)$.}) and it is
easy to distribute, making a good candidate for huge networks. Indeed, most
of the time is spent on sampling values of $\gamma$ to produce base
clusterings,\footnote{The combination of clusterings is extremely fast, as it is
linear in the number of nodes, rather than in the number of arcs, and has little
impact on the overall run time.} and this operation can be performed for each
$\gamma$ in a fully parallel way. Applying LLP to a web graph with 1
billion nodes and 50 billions arcs would require few hours in this setting.

For comparison, we also tried to compress our dataset using the alternative
versions of LLP described in Section~\ref{sec:lpa}: in particular, we
considered APM (with the optimal choice of $\gamma$) and the combination APM+Gray (that
sorts each APM cluster using Gray). Besides the number of bits per link, we also analysed
two measures that quantify two different structural properties:
\begin{itemize} 
	\item the \emph{average gap cost} (i.e.,~the average
of the base-2 logarithms of the gaps between the successors of a node: this is
an approximate measure of the number of bits required to write the gaps using a
universal variable-length code); this measure is intended to account for
locality: the average gap cost is small if the ordering tends to keep
well-connected nodes close to one another;\footnote{We remark that the average
gap cost is essentially an amortised version of the standard \emph{gap measure}
used in the context of data-aware compressed data structures~\cite{GHSCDS}.}
	\item the \emph{percentage of copied arcs} (i.e.,~the number of arcs that are
not written explicitly but rather obtained by reference from a previous
successor list); this is intended to account for similarity: this percentage is
small if the ordering tends to keep nodes with similar successor lists close to
one another.
\end{itemize} 
The results obtained are presented in Table~\ref{tab:llpavsapm}.
In most cases APM copies a smaller percentage of arcs than
APM+Gray, because Gray
precisely aims at optimising similarity rather than locality; this phenomenon is less pronounced on web graphs, where
anyway the overall number of copied arcs is larger;
looking at the average gap cost, all clustering methods turn out to do a better
job than Gray in improving locality (data not shown in the table). LLP
usually copies less arcs than APM+Gray, but the difference is often negligible and definitely balanced by the gain in
locality.

\begin{figure*}
\includegraphics[scale=.42]{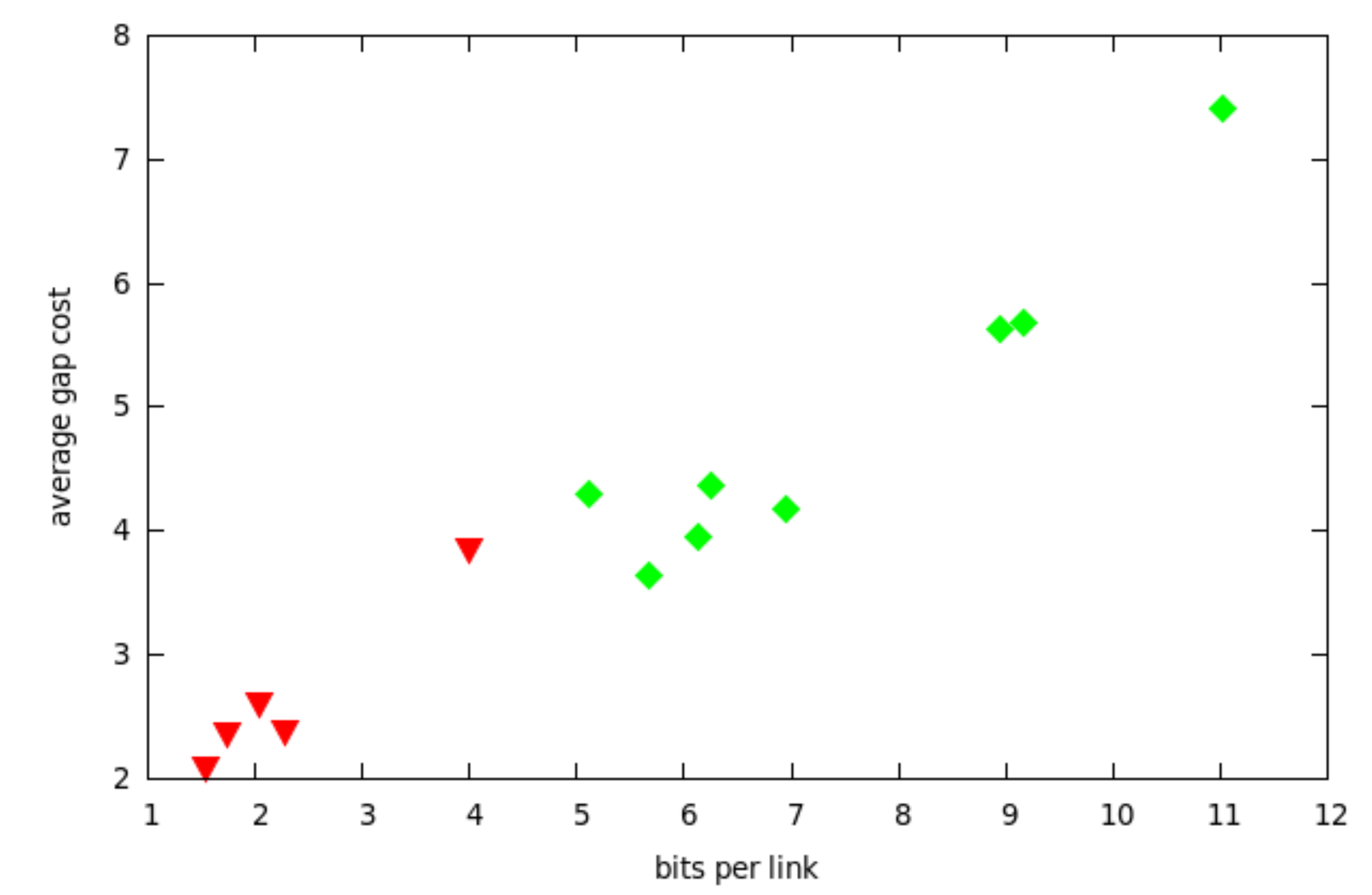}\includegraphics[scale=.42]{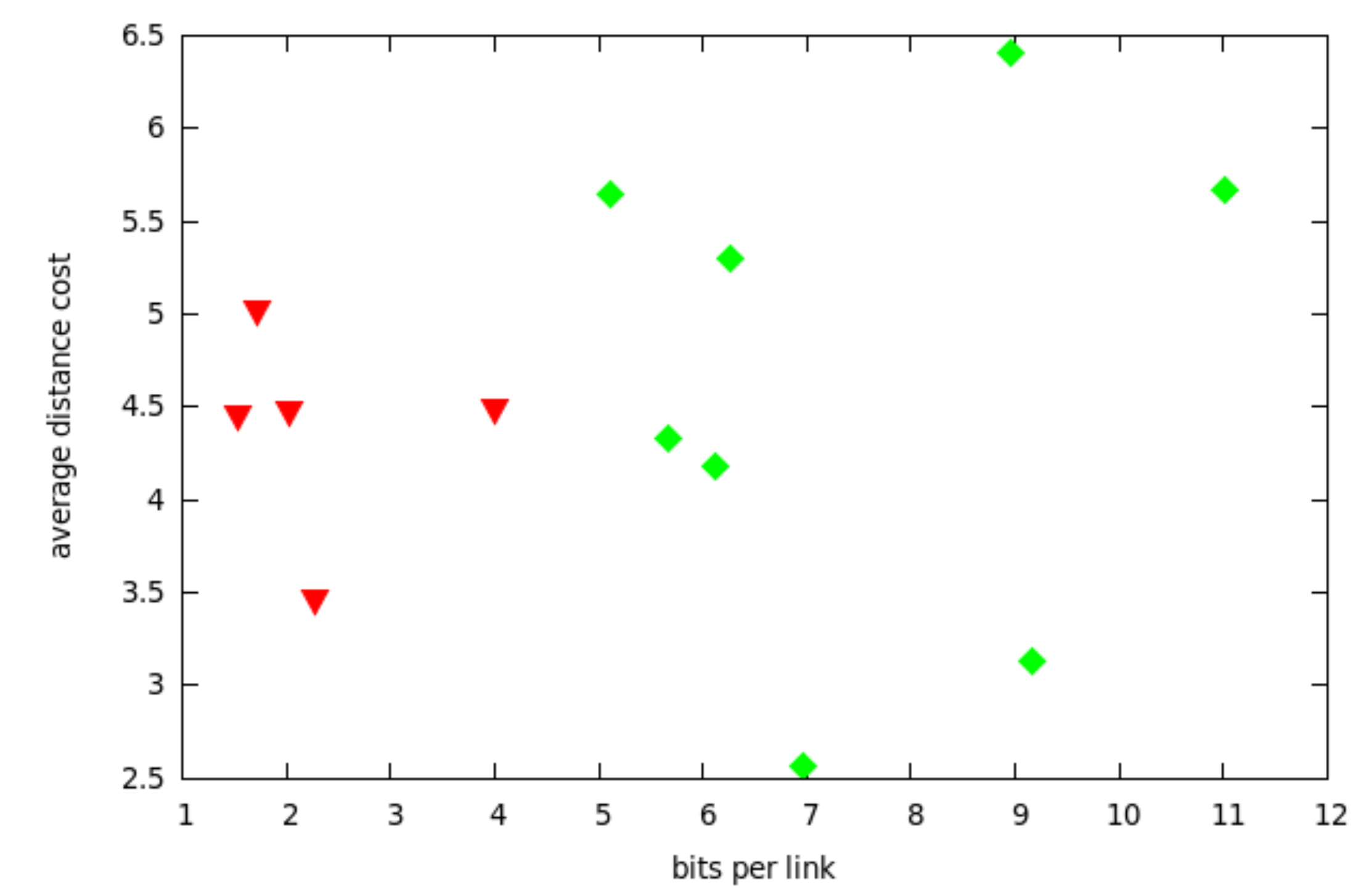}
\caption{\label{fig:bplgapdelta}Bits per link against average gap (left) and
distance (right) cost. $\bigtriangledown$ points indicates web graphs while $\Diamond$
points indicates social graph.}
\end{figure*}

We would like to point out that, at least when using the best compression
currently available (LLP+BV), the average gap cost is definitely more
correlated with compression rates than the \emph{average distance cost}, that is,
the average of the logarithms of the (absolute) difference between
the source and target of each arc (see Figure \ref{fig:bplgapdelta}). Indeed, the correlation coefficient is $0.9681$ between bits
per link and average gap cost and $0.1742$ between bits per link and average
distance cost. In~\cite{CKLCSN} the problems \textsc{MLogA} and \textsc{MLogGapA}
consist exactly in minimising the average distance and the average gap cost,
respectively: that authors claim that both problems capture the essence of a
good ordering, but our extensive experimentation suggests otherwise.

As a final remark, it is worth noticing that similarity and locality have a
different impact in social networks than in web graphs: in web graphs the
percentage of copied arcs is much larger (a clue of the presence of a better-defined structure) and in fact
it completely determines the number of bits per link, 
whilst in social networks the compression ratio is entirely established by the
gain of locality (measured, as usual, by the average gap cost).

\begin{table*}[t]
	\begin{center}
		
\begin{tabular}{ l r r r r r r r r r r r r}
\toprule
\textbf{Name} &\multicolumn{2}{c}{\textbf{LLP}} & \multicolumn{2}{c}{\textbf{BFS}} & \multicolumn{2}{c}{\textbf{Shingle}} & \multicolumn{2}{c}{\textbf{Gray}} & \multicolumn{2}{c}{\textbf{Natural}} & \multicolumn{2}{c}{\textbf{Random}} \\ \midrule
\texttt{Amazon} & 9.12 & (-30\%) & 13.01 & & 14.36 & (+10\%) & 13.11 & (+0\%) & 16.92 & (+30\%) & 23.62 & (+81\%) \\
\texttt{DBLP} & 6.87 & (-24\%) & 8.98 & & 11.39 & (+26\%) & 8.50 & (-6\%) & 11.36 & (+26\%) & 22.07 & (+145\%) \\
\texttt{Enron} & 6.45 & (-26\%) & 8.68 & & 9.80 & (+12\%) & 9.78 & (+12\%) & 13.43 & (+54\%) & 14.02 & (+61\%) \\
\texttt{Hollywood} & 5.17 & (-33\%) & 7.64 & & 6.68 & (-13\%) & 6.35 & (-17\%) & 15.20 & (+98\%) & 16.23 & (+112\%) \\
\texttt{LiveJournal} & 10.95 & (-28\%) & 15.05 & & 15.66 & (+4\%) & 14.19 & (-6\%) & 14.35 & (-5\%) & 23.50 & (+56\%) \\
\texttt{Flickr} & 8.90 & (-19\%) & 10.92 & & 10.22 & (-7\%) & 10.82 & (-1\%) & 13.87 & (+27\%) & 14.49 & (+32\%) \\
\texttt{indochina (hosts)} & 5.57 & (-15\%) & 6.55 & & 7.15 & (+9\%) & 7.49 & (+14\%) & 9.26 & (+41\%) & 10.59 & (+61\%) \\
\texttt{uk (hosts)} & 6.35 & (-17\%) & 7.59 & & 8.07 & (+6\%) & 8.13 & (+7\%) & 10.81 & (+42\%) & 15.58 & (+105\%) \\
\midrule
\texttt{eu} & 3.88 & (-21\%) & 4.87 & & 6.09 & (+25\%) & 4.98 & (+2\%) & 5.24 & (+7\%) & 19.89 & (+308\%) \\
\texttt{in} & 2.44 & (-26\%) & 3.29 & & 4.19 & (+27\%) & 2.90 & (-12\%) & 2.99 & (-10\%) & 21.15 & (+542\%) \\
\texttt{indochina} & 1.68 & (-24\%) & 2.21 & & 2.91 & (+31\%) & 2.12 & (-5\%) & 2.19 & (-1\%) & 21.46 & (+871\%) \\
\texttt{it} & 2.05 & (-26\%) & 2.76 & & 3.61 & (+30\%) & 2.67 & (-4\%) & 2.83 & (+2\%) & 26.40 & (+856\%) \\
\texttt{uk} & 1.80 & (-26\%) & 2.43 & & 3.26 & (+34\%) & 2.47 & (+1\%) & 2.75 & (+13\%) & 27.55 & (+1033\%) \\
\texttt{altavista-nd} & 5.25 & (-10\%) & 5.78 & & 8.12 & (+40\%) & 6.40 & (+10\%) & 8.37 & (+44\%) & 34.76 & (+501\%) \\
\bottomrule
\end{tabular}

	\end{center}
	\caption{Compression results starting from natural order
	(percentages are relative to BFS). Values are bits per link.}
	\label{tab:bits_nonrandom}
\end{table*}

\begin{table*}[t]
	\begin{center}
		
\begin{tabular}{ l r r r r r r r r r r r r}
\toprule
\textbf{Name} &\multicolumn{2}{c}{\textbf{LLP}} & \multicolumn{2}{c}{\textbf{BFS}} & \multicolumn{2}{c}{\textbf{Shingle}} & \multicolumn{2}{c}{\textbf{Gray}} & \multicolumn{2}{c}{\textbf{Natural}} & \multicolumn{2}{c}{\textbf{Random}} \\ \midrule
\texttt{Amazon} & 9.16 & (-30\%) & 12.96 & & 14.43 & (+11\%) & 14.60 & (+12\%) & 16.92 & (+30\%) & 23.62 & (+82\%) \\
\texttt{DBLP} & 6.88 & (-23\%) & 8.91 & & 11.42 & (+28\%) & 11.50 & (+29\%) & 11.36 & (+27\%) & 22.07 & (+147\%) \\
\texttt{Enron} & 6.51 & (-24\%) & 8.54 & & 9.87 & (+15\%) & 9.94 & (+16\%) & 13.43 & (+57\%) & 14.02 & (+64\%) \\
\texttt{Hollywood} & 5.14 & (-35\%) & 7.81 & & 6.72 & (-14\%) & 6.40 & (-19\%) & 15.20 & (+94\%) & 16.23 & (+107\%) \\
\texttt{LiveJournal} & 10.90 & (-28\%) & 15.1 & & 15.77 & (+4\%) & 15.73 & (+4\%) & 14.35 & (-5\%) & 23.50 & (+55\%) \\
\texttt{Flickr} & 8.89 & (-22\%) & 11.26 & & 10.22 & (-10\%) & 10.23 & (-10\%) & 13.87 & (+23\%) & 14.49 & (+28\%) \\
\texttt{indochina (hosts)} & 5.53 & (-17\%) & 6.63 & & 7.16 & (+7\%) & 7.49 & (+12\%) & 9.26 & (+39\%) & 10.59 & (+59\%) \\
\texttt{uk (hosts)} & 6.26 & (-18\%) & 7.62 & & 8.12 & (+6\%) & 8.13 & (+6\%) & 10.81 & (+41\%) & 15.58 & (+104\%) \\
\midrule
\texttt{eu} & 3.90 & (-21\%) & 4.93 & & 6.86 & (+39\%) & 6.27 & (+27\%) & 5.24 & (+6\%) & 19.89 & (+303\%) \\
\texttt{in} & 2.46 & (-30\%) & 3.51 & & 4.79 & (+36\%) & 4.40 & (+25\%) & 2.99 & (-15\%) & 21.15 & (+502\%) \\
\texttt{indochina} & 1.71 & (-26\%) & 2.31 & & 3.59 & (+55\%) & 3.09 & (+33\%) & 2.19 & (-6\%) & 21.46 & (+829\%) \\
\texttt{it} & 2.10 & (-28\%) & 2.89 & & 4.39 & (+51\%) & 3.79 & (+31\%) & 2.83 & (-3\%) & 26.40 & (+813\%) \\
\texttt{uk} & 1.91 & (-33\%) & 2.84 & & 4.09 & (+44\%) & 3.36 & (+18\%) & 2.75 & (-4\%) & 27.55 & (+870\%) \\
\texttt{altavista-nd} & 5.22 & (-11\%) & 5.85 & & 8.12 & (+38\%) & 7.52 & (+28\%) & 8.37 & (+43\%) & 34.76 & (+494\%) \\
\bottomrule
\end{tabular}

	\end{center}
	\caption{Compression results starting from a random order (percentages
	are relative to BFS). Values are bits per link.}

	\label{tab:bits_random}
\end{table*}

\begin{table*}[t]
	\begin{center}
		
\begin{tabular}{ l ccc|ccc|ccc }
\toprule
\textbf{Name} & \multicolumn{3}{c}{\textbf{Bits/link}}&\multicolumn{3}{c}{\textbf{Copied arcs}}&\multicolumn{3}{c}{\textbf{Avg.~gap cost}} \\
& \textbf{LLP}&\textbf{APM + Gray}&\textbf{APM}&\textbf{LLP}&\textbf{APM + Gray}&\textbf{APM}&\textbf{LLP}&\textbf{APM + Gray}&\textbf{APM} \\ \midrule
\texttt{Amazon} &	 9.14&	10.45&	10.67&	31.22&	32.32&	28.87&	5.64&	6.87&	6.97 \\
\texttt{DBLP} &		 6.87&	8.38&	8.48&	36.55&	37.66&	36.42&	4.04&	5.73&	5.80 \\
\texttt{Enron} &	 6.48&	7.15&	7.97&	24.07&	25.45&	10.86&	3.92&	4.58&	4.76 \\
\texttt{Hollywood} &	 5.13&	5.38&	6.10&	44.22&	42.49&	38.68&	4.14&	4.38&	4.92 \\
\texttt{LiveJournal} &	 10.90&	12.00&	12.79&	23.57&	23.66&	17.48&	7.34&	8.29&	8.69 \\
\texttt{Flickr} &	 8.89&	9.22&	9.69&	13.65&	11.77&	8.88&	5.59&	5.84&	6.17 \\
\midrule
\texttt{eu} &			3.90&	4.86&	5.76&	65.84&	66.33&	59.57&	3.62&	5.16&	5.78 \\
\texttt{in} &	 		2.46&	3.11&	4.05&	72.45&	73.04&	65.11&	2.31&	4.02&	4.60 \\
\texttt{indochina} &		1.71&	2.17&	3.00&	80.36&	80.78&	75.53&	2.06&	3.59&	4.09 \\
\texttt{indochina (hosts)} &	5.54&	6.04&	6.16&	33.51&	34.69&	26.94&	3.46&	4.02&	3.90 \\
\texttt{it} &			2.10&	2.56&	3.94&	77.18&	79.76&	69.53&	2.43&	4.36&	5.16 \\
\texttt{uk (hosts)} &	 	6.26&	6.68&	6.90&	33.94&	37.34&	30.48&	4.24&	4.76&	4.79 \\
\texttt{uk} &		 	1.91&	2.39&	3.73&	79.16&	81.92&	71.73&	2.31&	4.71&	5.35 \\
\bottomrule
\end{tabular}

	\end{center}
	\caption{Comparison between LLP and the ordering produced by other
	clustering algorithms (APM and the combination APM+Gray) when compressing with
	the BV algorithm. We consider the value of $\gamma$ that minimises the number
	of bits/link.}
	\label{tab:llpavsapm}
\end{table*}

\section{Conclusions and Future Work}

We have presented highly scalable techniques that improve compressed data
structures for representing web graphs and social networks significantly beyond
the current state-of-art. More importantly, we have shown that coordinate-free
methods can outperform state-of-art extrinsic techniques on a large range of
networks. The clustering techniques we have devised are scalable to billions of
nodes, as they just require few linear passes over the graphs involved. In some
cases (e.g., the \texttt{uk} dataset) we bring down the cost of a link to 
$1.8$ bits. We remark again that our improvements are measured w.r.t.~the BFS
baseline, which is itself often an improvement when compared to the existing
literature.

Finally, we leave for future work a full investigation of the compression ratio
that can be obtained when fast access is not required. For instance, \texttt{uk}
compressed by LLP+BV at maximum compression requires only $1.21$ bits per
link---better, for instance, than the Apostolico--Drovandi method with maximum
compression ($1.44$). Some partial experimental data suggests
that we would obtain by far the highest compression ratio currently available.

The experiments that we report required several thousands of hours of
computation: we plan to make available the results both under the form of
WebGraph \emph{property files} (which contain a wealth of statistical data) and
under the form of comprehensive graphical representations. 


\bibliography{law}

\end{document}